\documentclass[doublecol]{epl2} 
\usepackage{braket}

\begin{document}

\title{Dynamical thermalization \\ in isolated quantum dots and black holes}

\author{Andrey R. Kolovsky\inst{1,2} \and  Dima L.Shepelyansky\inst{3}} 
\shortauthor{A.R.Kolovsky, D.L.Shepelyansky}
%\shorttitle{Dynamical thermalization of isolated quantum dots and black holes}
\institute{
 \inst{1}  Kirensky Institute of Physics, 660036 Krasnoyarsk, Russia\\
 \inst{2} Siberian Federal University, 660041 Krasnoyarsk, Russia\\
 \inst{3} Laboratoire de Physique Th\'eorique du CNRS (IRSAMC), 
  Universit\'e de Toulouse, UPS, F-31062 Toulouse, France
}
%\date{\today} 
%\date{December XX, 2016}

\abstract{We study numerically a model of 
quantum dot with interacting fermions.
At strong interactions with small conductance
the model is reduced to the Sachdev-Ye-Kitaev black hole model
while at weak interactions and large conductance
it describes a Landau Fermi liquid in a regime of 
quantum chaos. We show that above the {\AA}berg threshold
for interactions there is an onset of dynamical themalization
with the Fermi-Dirac distribution describing the eigenstates of
isolated dot. At strong interactions in the isolated black hole regime
there is also onset of dynamical thermalization with the
entropy described by the quantum Gibbs distribution.
This dynamical thermalization takes place in an isolated system without
any contact with thermostat.
We discuss possible realization of these regimes 
with quantum dots of 2D electrons and cold ions in optical lattices.  
}


%\date{December XX, 2016}
%\date{\today}

\maketitle

\section{Introduction}

Recently it has been shown that there is a duality relation between 
an isolated quantum dot with 
infinite-range strongly interacting  
fermions and  a quantum Black Hole model in
$1+1$ dimensions \cite{sachdev1993,kitaev2015,sachdev2015}.
This system, called the Sachdev-Ye-Kitaev (SYK) model,
attracted a significant interest of quantum gravity
community (see e.g. \cite{polchinski,maldacena1,stanford,gross}).
A possible realization of SYK model with ultracold atoms
in optical lattices has been proposed recently in \cite{hanada}.
An important element of the SYK model is an emergence of 
many-body quantum chaos with a maximal 
Lyapunov exponent for dynamics in a semiclassical limit
\cite{maldacena1,maldacena2}.

It should be noted that in fact the SYK model
first appeared in the context of 
nuclear physics 
where the strongly interacting fermions
had been described by the two-body random interaction 
model (TBRIM) with all random two-body matrix elements
between fermions on degenerate energy orbitals
\cite{bohigas1,bohigas2,french1,french2}.
The majority of matrix elements of the TBRIM
Hamiltonian is zero since the two-body interactions
impose selective transition rules. Thus this case is rather
different from the case of Random Matrix Theory (RMT)
introduced by Wigner for description of complex nuclei, atoms and molecules
\cite{wigner,mehta}. In spite of this difference, it had been shown
that the level spacing statistics $P(s)$ in TBRIM is described
by the Wigner-Dyson distribution  $P_W(s)$ typical for the RMT
\cite{bohigas2,french2}. A similar situation appeared later
for the models of quantum chaos also characterized
by sparse matrices being rather far from the RMT type
but also characterized by the  Wigner-Dyson statistics
for matrices of a specific symmetry class \cite{bohigas,haake,ullmo}. 
The validity of RMT for the SYK model at different symmetries
has been demonstrated in the recent fundamental and detailed
studies reported in \cite{garcia}. Transport properties of SYK and 
its extensions are also discussed recently \cite{ageorges}. 

In this Letter we present the studies of SYK type model 
extending parallels between  physics of quantum dots and 
black holes. Indeed, the TBRIM and SYK systems
have degenerate non-interacting orbitals
while for quantum dots it is natural to 
have nondegenerate orbitals characterized by a certain
average one-particle level spacing $\Delta$. This spacing $\Delta$
can be much smaller than a typical interaction strength $U$
between fermions and then we have the usual SYK or TBRIM degenerate
regime ($\Delta \ll U$). In the opposite limit 
we have the regime of weak interactions
typical for metallic dots ($\Delta \gg U \approx \Delta/g$) 
where the interaction is determined by a dimensionless dot conductance  
$g=E_c/\Delta$ with $E_c$ being the Thouless energy \cite{thouless,akkermans}.
Even if the average spacing between exited levels drops
exponentially with the number of fermions the mixing
of levels takes place only at interactions $U$ being
much larger than this spacing since the two-body selection rules
connect directly only a polynomial number of states.
The border for emergence of the RMT $P_W(s)$ statistics
is determined by the  {\AA}berg criterion  \cite{aberg1,aberg2}
telling that the transition to RMT takes place when the 
average two-body matrix elements become larger than the 
average spacing between directly coupled states.
This criterion has been confirmed in extensive numerical simulations 
with interacting fermions and spin systems 
\cite{jacquod,georgeot,benenti,dlsnobel}. 
Thus at $g \gg 1$
the RMT statistics appears only for relatively high excitation above
the quantum dot Fermi energy $E_F$ \cite{aberg1,aberg2,jacquod}:
\begin{equation}
\label{eq1}
\delta E = E - E_F > \delta E_{ch} \approx g^{2/3} \Delta \;\; .
\end{equation}
This border is in a good agreement with the spectroscopy
experiments of individual mesoscopic quantum dots \cite{sivan}. 
Of course, an exact check of  the  {\AA}berg criterion 
via numerical simulations is not an easy task since the matrix size grows 
exponentially with the number of fermions. Thus the validity
of the {\AA}berg border (\ref{eq1}) is still under active 
discussions in relation to
the Many-Body Localization-delocalization (MBL) transition
(see \cite{polyakov} and Refs. therein). We also note that 
the quantum chaos and RMT statistics in 
the interacting Bose systems has been studied 
in \cite{kolovsky}.  

In fact the border (\ref{eq1}) assumes also that
the emergence of RMT statistics appears as a result of
onset of quantum ergodicity which in its turn leads to
a dynamical thermalization in an isolated system \cite{jacquod}.
Thus the relation (\ref{eq1}) is based on a rather general 
Dynamical Thermalization Conjecture (DTC). 
According to the DTC individual eigenstates of isolated system
are described by the standard Fermi-Dirac thermal 
distribution. The examples of thermalized
individual eigenstates have been presented in \cite{benenti}.
This individual eigenstate thermalization 
is more striking than the thermal distribution of
probabilities averaged over a group of eigenstates
which had been seen earlier in the numerical simulations
of TBRIM with $\Delta > U$ \cite{flambaum}.
At present, the dynamical thermalization of individual
eigenstates is known as the Eigenstate Thermalization Hypothesis
(ETH) and attracts a significant interest of the scientific community
%both in theory and experimental studies with cold atoms
(see e.g.  \cite{huse,polkovnikov,borgonovi}).

The direct check of DTC from the filling factors
of one-particle orbitals is possible 
but it requires  diagonilization of large matrices  
and still the fluctuations are significant even for sizes 
$N \sim 10^7$ (see e.g. Fig.6 in \cite{benenti}).
In fact it has been found that the fluctuations are
significantly reduced if we determine numerically 
the dependence of entropy $S$ on energy $E$
computed for  individual eigenstates.
The reduction of fluctuations is due to the fact
that both $S$ and $E$ are extensive self-averaging
characteristics. The numerically obtained dependence $S(E)$
can be directly compared with those of the theoretical
Fermi-Dirac or Bose-Einstein distributions
\cite{landau}. The power of this approach
for a verification of DTC has been confirmed in
the numerical simulations of classical 
nonlinear disordered chains \cite{ermannnjp},
the  Gross-Pitaevski equation for Bose-Einstein
condensate in chaotic billiards \cite{stadium,sinaiosc}
and quantum many-body Bose-Hubbard rings with disorder \cite{schlagheck}.
Here we use this approach for investigation
of dynamical thermalization in isolated quantum dots and 
black holes described by TBRIM and SYK type models.
 
\section{Model description}

The model is described by the Hamiltonian
for $L$ spin-polarized fermions on $M$ energy orbitals $\epsilon_k$ 
($\epsilon_{k+1} \geq \epsilon_k$):
\begin{eqnarray}
\label{eq2}
\nonumber
\widehat{H} &=& \widehat{H}_0 + \widehat{H}_{int} \;, 
\; \widehat{H}_0 = \frac{1}{\sqrt{M}} 
\sum_{k=1}^M v_k \hat{c}_k^\dagger \hat{c}_k  \;, \\
\nonumber \\
\widehat{H}_{int} &=&  \frac{1}{\sqrt{2M^3}} \sum_{ijkl} J_{ij,kl}  
\hat{c}_i^\dagger  \hat{c}_j^\dagger   \hat{c}_k \hat{c}_l \;. 
\end{eqnarray} 
Here the fermion operators $\hat{c}_i^\dagger, \hat{c}_i$ satisfy
the usual anti-commutation relation. The interaction matrix elements
$J_{ij,kl} $ are random complex variables with a standard deviation $J$
and zero average value.
The interacting part $\widehat{H}_{int}$ is the same as those 
used in \cite{hanada} (see Eq.(1) there). As in \cite{hanada} we 
consider the model with complex fermions (complex matrix elements
$J_{ij,kl}$) \cite{sachdev2015}
which is slightly different from the case with real fermions 
(real $J_{ij,kl}$) \cite{kitaev2015}.
However, in our model (\ref{eq2}), in addition to the interaction Hamiltonian
$\widehat{H}_{int}$, there is also the unperturbed
part $\widehat{H}_0$ describing one-particle orbitals $\epsilon_k=v_k/\sqrt{M}$
in a quantum dot of non-interacting fermions.
The average of one-orbital energies is taken to be
$\overline{v_k^2}=V^2$ with
$\overline{v_k}=0$.
Thus the unperturbed one-particle energies $\epsilon_k$ are distributed in
an energy band of size $V$ and the 
average level spacing between them is $\Delta \approx V/M^{3/2}$ 
while the two-body coupling matrix element is
$U \approx J/M^{3/2}$. Hence, in our model  the effective dimensionless
conductance in (\ref{eq1}) is $g = \Delta /U \approx  V/J$.
The total matrix size of the Hamiltonian (\ref{eq2}) is
$N={M!}/{L!(M-L)!}$ and each multi-particle state is coupled
with $K=1+L(M-L)+L(L-1)(M-L)(M-L-1)/4$ states \cite{jacquod,flambaum,borgonovi}.
Here we consider the case of approximate half filling with $L \approx M/2$.

\section{Dynamical thermalization ansatz}

We start from the case of $g \gg 1$ when one-particle orbitals
are well defined. If the DTC is valid, then 
weak or moderate interactions should lead to the standard
Fermi-Dirac thermal distribution over $M$ one-particle orbitals
with energies $\epsilon_k$ and filling factors \cite{landau}:
\begin{equation}
\label{eq3}
{n}_k = \frac{1}{e^{\beta(\epsilon_k - \mu)} + 1}  \;, \;\; \beta = 1/T \;, 
\end{equation}
with the chemical potential $\mu$ determined by the
conservation of number of fermions
$\sum_{k=1}^{M} {n}_k =L$. Then following \cite{landau},  
at a given temperature
$T$, the system energy $E$ and  entropy $S$
are given  by 
\begin{equation}
\nonumber
\label{eq4}
E(T)= \sum_{k=1}^M \epsilon_k {n}_k \;,\quad
S(T) = - \sum_{k=1}^{M} {n}_k  \ln {n}_k \; .
\end{equation}
These two relations determine an implicit
functional dependence $S(E)$ which is very convenient
for numerical checks since both quantities
$S$ and $E$ are extensive and self-averaging.
The numerical computation of $S$ and $E$
from the eigenstates $\psi_m$ 
and eigenenergies $E_m$ of $H$ is straightforward
by using 
$n_k(m)=\braket{\psi_m|\hat{c}_k^\dagger \hat{c}_k|\psi_m}$.
With (\ref{eq4}) this gives
entropy $S_m$ of eigenstate $\psi_m$.

%As usual we also have $\beta=1/T = \partial S/\partial E$ \cite{landau}.

\begin{figure}[h]
\begin{center}
\includegraphics*[width=8.2cm]{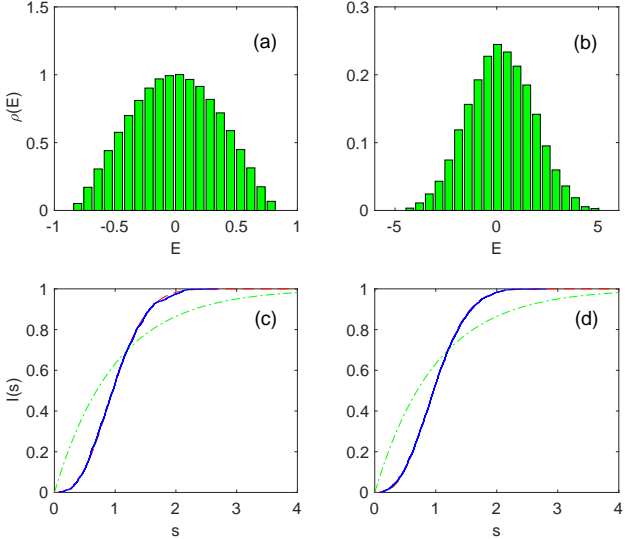}
\end{center}
\vglue -0.3cm
\caption{(Color on-line) 
Top row: density of states %in energy 
$\rho(E) = d N(E)/d E$
for the model (\ref{eq2}). Bottom row (c,d):
integrated level spacing statistics $I(s) =\int_0^s ds' P(s')$
for the Poisson statistics $P_P(s)$ (dashed green curve),
for the Wigner surmise $P_W(s)$ (dashed red curve)
and numerical data $P(s)$ 
for central energy region comprising 80 percent of the states
(blue curve, almost superposed with red dashed curve). 
Here $M=14, L=6, N=3003$, and $J=1, V=0$ (a,c)
and $J=1, V=\sqrt{14}$ (b,d).
}  
\label{fig1}
\end{figure}

Of course, the DTC is based on a quantum ergodicity of eigenstates
that can appear only in the regime of Wigner-Dyson statistics
$P_W(s) = 32 s^2 \exp(-4s^2/\pi)/\pi^2$ 
(Wigner surmise corresponding to the Gaussian Unitary Ensemble (GUE) symmetry
of our model). Indeed,
in absence of ergodicity the statistics 
is describes by the Poisson
distribution $P_P(s) = \exp(-s)$ of independent uncorrelated 
eigenenergies.
As usual, here the level spacing $s$, 
between adjacent eigenenergies $E_m, E_{m+1}$ of the whole system, 
is measured in units of
average level spacing assuming spectrum unfolding 
\cite{haake,ullmo}). 

Indeed, our results presented in Fig.~\ref{fig1},
show that the RMT is valid for the SYK black hole regime
($g=V/J \ll 1$) and for the quantum dot regime
($g=V/J > 1$) when the relation (\ref{eq1}) 
is satisfied. We note that the Wigner-Dyson statistics 
at $V=0$ had been also obtained in \cite{bohigas2,french2,garcia}
for corresponding symmetries.

\begin{figure}[t]
\begin{center}
\includegraphics*[width=8.2cm]{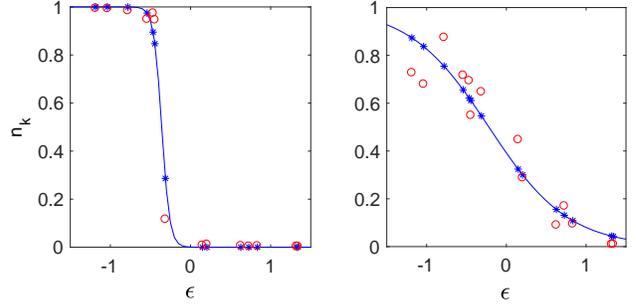}
\end{center}
\vglue -0.3cm
\caption{(Color on-line) 
Dependence of filling factors $n_k$
on energy $\epsilon$ for individual eigenstates
obtained from exact diagonatization of (\ref{eq2})
(red circles) and from the Fermi-Dirac ansatz 
with one-particle energy $\epsilon$ (\ref{eq3})
(full blue curve; blue stars 
are shown at one-particle energy positions
$\epsilon=\epsilon_k$). Here
$M=14, L=6, N=3003$, $J=1, V=\sqrt{14}$ and 
eigenenergies (\ref{eq2}) are $E=-4.4160$ (left panel), $-3.0744$ (right panel);
the theory blue curves (\ref{eq3}) are drown for the
temperatures corresponding to these energies
$\beta=1/T=20$ (left panel), 
$2$ (right panel), see Fig.~\ref{fig3}(b) below.
}  
\label{fig2}
\end{figure}

\begin{figure}[t]
\begin{center}
\includegraphics*[width=8.2cm]{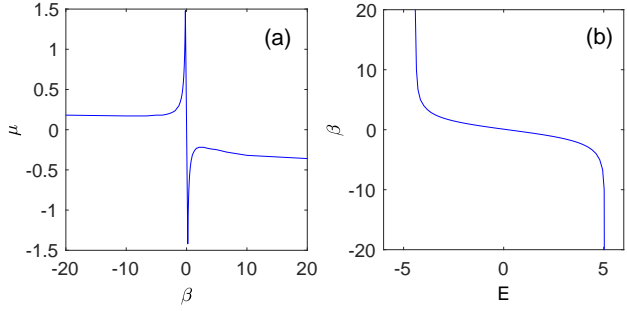}
\end{center}
\vglue -0.3cm
\caption{(Color on-line) 
Dependence of inverse temperature $\beta=1/T$ on energy $E$
(right panel) and chemical potential $\mu$ on $\beta$ (left panel) 
given by the Fermi-Dirac ansatz (\ref{eq3})
for the set of one-particle energies $\epsilon_k$ as in  Fig.~\ref{fig2}.
}  
\label{fig3}
\end{figure}

For the parameters of Fig.~\ref{fig1} (right column)
we indeed find that the DTC provides a good description
of filling factors in agreement with (\ref{eq3})
as it is shown for two specific eigenstates in Fig.~\ref{fig2}.
 However, the fluctuations are
significant and also the DTC should be verified for all
eigenstates at a given set of parameters.
Due to that we test the DTC validity using the approach
developed for bosons in \cite{stadium,sinaiosc,schlagheck}
based on the numerical computation of the dependence $S(E)$
from eigenstates of Hamiltonian (\ref{eq2}).

\section{Quantum dot regime}

First of all we remark that,
since the energy spectrum of our system (\ref{eq2}) 
is inside a finite energy band, it is possible to have
also negative temperatures for energies being in the upper half
of energy band. Such a regime of negative temperatures
is well known for spin systems \cite{abragam}. 
The relation between the temperature $T$ and the energy $E$
and the dependence of chemical potential $\mu$
on $T$, obtained from the Fermi-Dirac ansatz (\ref{eq3}),
are shown in Fig.~\ref{fig3}.
The center of the energy band at $E=0$ corresponds to infinite
temperature and $\beta=0$. Here the entropy takes its maximal value
$S(E=0) = -L \ln(L/M)$ corresponding to equipartition of 
$L$ fermions  over $M$ orbitals. With increase of $|\beta|$ the entropy 
obviously decreases towards zero, which corresponds to unit filling factor for 
$L$ lowest ($T =+0$) or highest ($T=-0$) energy orbitals. 

\begin{figure}[h]
\begin{center}
\includegraphics*[width=8.2cm]{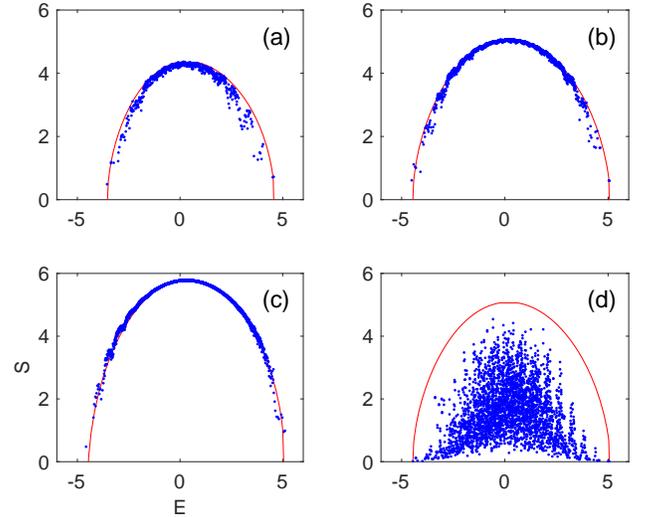}
\end{center}
\vglue -0.3cm
\caption{(Color on-line) 
Dependence of entropy on energy $S(E)$ 
for (a) $M=12$, $L=5$, $N=792$, $J=1$;
(b) $M=16$, $L=7$, $N=3003$, $J=1$;
(c) $M=14$, $L=6$, $N=11440$, $J=1$;
(d) $M=16$, $L=7$, $N=3003$, $J=0.1$.
Blue points show the numerical data $E_m, S_m$
for all eigenstates, the red curves show
the theoretical Fermi-Dirac thermal distribution
(\ref{eq3}). Here $V=\sqrt{14}$. 
}  
\label{fig4}
\end{figure}

The dependence $S(E)$ for DTC in the quantum dot regime
at $g = V/J > 1 1$, is shown 
in Fig.~\ref{fig4} (a,b,c). Here the conductance of the dot
is not very large ($g =V/J \approx   3.7$)
and practically all eigenstates are well thermalized
with numerical points following the DTC theoretical curve.
This is in agreement with the estimate (\ref{eq1})
which gives the thermalization at rather low energy excitation
$\delta E \approx 0.17$. In contrast, for $J=0.1$ ((d) panel)
we have significant increase of $g=37$ with larger
border for the RMT statistics $\delta E \approx 0.8$.
As a result  the numerical data have entropy $S$ 
significantly below the theoretical value.

\begin{figure}[h]
\begin{center}
\includegraphics*[width=8.2cm]{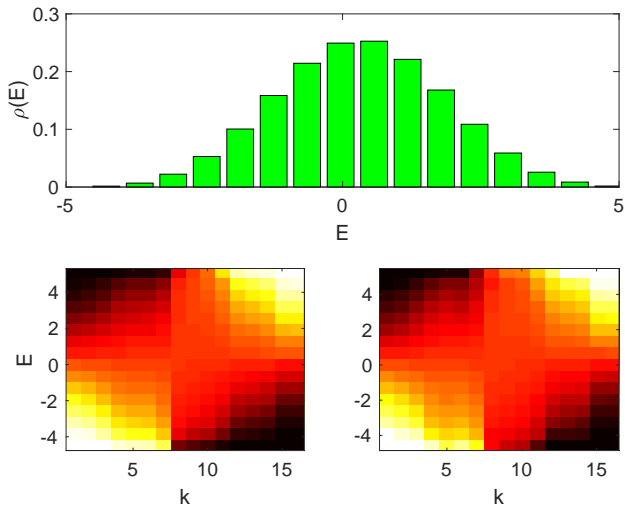}
\end{center}
\vglue -0.3cm
\caption{(Color on-line) 
Top panel: dependence of normalized density of states $\rho(E)$
on energy $E$ ($\int  \rho(E) dE = 1$),
$\rho(E)$ averaged inside each energy cell.
Bottom panels: occupation probabilities 
$n_k(E)$ of one-particle orbitals $\epsilon_k$
given by the theoretical Fermi-Dirac distribution (\ref{eq3})
on left panel, and by their numerical values
obtained by exact diagonalization of (\ref{eq2})
on right panel; $n_k$ is averaged over all eigenstates
inside a given energy cell.
The color changes from black for $n_k=0$ via red, yellow
to white for $n_k=1$; orbital number $k$ and eigenenergy $E$
are shown on $x$ and $y$ axes respectively.
Here $M=16, L=7, N=11440, V=4, J=1$.   
}  
\label{fig5}
\end{figure}

For each eigenstate with eigenenergy $E$ it is possible to determine
the occupation probabilities $n_k(E)$ on one-particle orbitals 
with orbital energies $\epsilon_k$. In the DTC regime the
dependence $n_k(E)$ is given by the Fermi-Dirac distribution (\ref{eq3})
shown in Fig.~\ref{fig5} (bottom left panel). 
The numerically obtained values $n_k(E)$, shown in 
Fig.~\ref{fig5} (bottom right panel), demonstrate a good
agreement with  the theory (\ref{eq3}). This
confirms the validity of the DTC for 
practically all eigenstates in the quantum dots with moderate
values of conductance ($g=4$ in Fig.~\ref{fig5}).

Here we presented results for one specific disorder realization.
Similar results have been obtained for other disorder realizations.
However, we do not present averaging over disorder since
it is much more striking that
the dynamical thermalization takes place 
even for one specific quantum dot with a given
disorder realization.

\section{SYK black hole regime}
\begin{figure}[h]
\begin{center}
\includegraphics*[width=8.2cm]{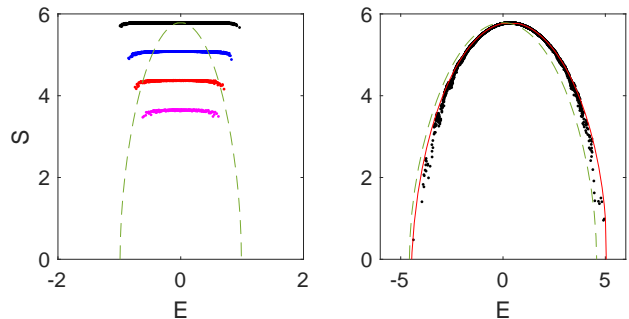}
\end{center}
\vglue -0.3cm
\caption{(Color on-line) 
Dependence $S(E)$ for the SYK black hole regime
at $V=0$
(left panel) and the quantum dot regime 
$V= \sqrt{14}$ (right panel).
Here $M=16, L=7, N=11440$ (black points),
$M=14, L=6, N=3003$ (blue points), 
$M=12, L=5, N=792$ (red points),
$M=10, L=4, N=210$ (magenta points); in all cases $J=1$.
Points show numerical data $E_m, S_m$
for all eigenstates, the full red curve shows the
theoretical Fermi-Dirac distribution 
(\ref{eq3}) in the right panel.
Dashed gray curves in both panels show
the Fermi-Dirac distribution (\ref{eq3})
for a semi-empirical model of non-interacting
quasi-particles for the case of black points (see text).
}  
\label{fig6}
\end{figure}

This regime corresponds to $g=V/J \ll 1$.
In this case we find rather different dependence
$S(E)$ shown in Fig.~\ref{fig6} (left panel) for several 
system sizes. In fact, here the entropy
has its maximal value $S = -L \ln (L/M) \approx L \ln 2 $
remaining practically independent of energy $E$
in a broad energy interval
in the center of energy band.
Only at the spectrum edges there is
a small decrease of entropy approximately by
10\% of its maximal value. 
Here, as before, the values of $S$ are obtained
from $n_k$ filling factors
computed numerically from 
eigenstates $\psi_m$ of (\ref{eq2}) by their projection 
on non-interacting orbitals of $H_0$.
The obtained dependence $S(E)$
is in a striking contrast with those
of the quantum dot regime shown in 
the right panel of Fig.~\ref{fig6}
for comparison.

The fact that in the SYK model the entropy
is practically independent of energy $E$ 
is not so surprising: the interactions are much
stronger than the energies of one-particle orbitals
so that many-body eigenstates are spread over all
orbitals giving for them almost constant filling factors $n_k$
and, hence, a constant entropy. 

The distinguished feature of the SYK model
is absence of quasi-particles with natural orbital energies so
that it is not possible to use the Fermi-Dirac ansatz (\ref{eq3})
which worked so well for the quantum dot regime.
Thus, to handle this case in the spirit of DTC
we assume that there are some hidden quasi-particles
which have certain one-particle orbitals $\epsilon_k$
and relatively weak interactions.
For unknown $\epsilon_k$ values we require
that the many-body density of states $\rho(E)$
found numerically at $V=0$ 
is reproduced by many-body 
eigenenergies of noninteracting fermions 
located on $\epsilon_k$ orbitals
(see e.g. Fig.~\ref{fig1}(a)).
As a first approximation we take equidistant
values $\epsilon_k$ in a certain interval
so that minimal and maximal energies
of many-body Hamiltonian (\ref{eq2}) at $V=0$
are $E_{min} = \sum_{k=1}^{L} \epsilon_k $
and $E_{max}=  \sum_{k=M-L}^{M}  \epsilon_k$
with equidistant $\epsilon_k$ values.
Then with these $\epsilon_k$ values
and the Fermi-Dirac ansatz (\ref{eq3})-(\ref{eq4})
we find that the many-body density of states $\rho(E)$,
obtained from the exact diagonalization of (\ref{eq2}),
is well reproduced. Such an approach can be applied
both with well defined quasi-particles ($g>1$)
and hidden quasi-particles $g \ll 1$).
The obtained $\rho(E)$ is not sensitive
to a randomization of $\epsilon_k$
at the fixed energy range $(E_{min},E_{max})$.

The obtained dependence $S(E)$
is shown in Fig.~\ref{fig6} (right panel)
for the quantum dot regime and we see that such a
semi-empirical dashed gray curve is rather close to the theoretical distribution
(\ref{eq3}) obtained with one-particle orbitals
at $g=3.74$. Thus we find that this semi-empirical 
approach works well in the regime $g>1$.
The semi-empirical results for the SYK black hole 
are shown in Fig.~\ref{fig6} (left panel, dashed gray curve). 
Here the semi-empirical curve 
correctly describes the maximal value of $S$ at $E \approx 0$
but is does not reproduce the large plateau obtained with numerical
data from $n_k=\braket{\psi_m|\hat{c}_k^\dagger \hat{c}_k|\psi_m}$
and (\ref{eq4}). We explain this difference
by the fact that $\hat{c}_k^\dagger , \hat{c}_k$
operators are written in the initial degenerate basis
with $\epsilon_k=v_k/\sqrt{M}=0$. This original basis
does not correspond to the basis and energies  of hidden
quasi-particles which are non-degenerate.
We expect that a certain linear transformation
can create a basis of new hidden quasi-particles
with the filling factors given the curve $S(E)$
being close to the semi-empirical curve in
Fig.~\ref{fig6} (left panel). However, the determination
of such a basis remains a further challenge.

For the SYK model we obtain numerically that
the entropy of the ground state is approximately
$S(T=0) \approx L \ln 2   \approx 0.69 L$ corresponding to
our approximate half filling $L/M \approx 0.5$.
This value is approximately by a factor $3$
larger than the numerical value $S(T=0)=0.21 L$
obtained in \cite{garcia} which is
close to the theoretical value $S(T=0)=0.23 L$
obtained in  \cite{maldacena1}.
We attribute this difference of numerical prefactor to
the fact that here we consider complex fermions
at an approximate half filling while
the Majorana fermions have been considered 
in \cite{garcia,maldacena1}.
We note that the maximal entropy value,
obtained in \cite{hanada} for complex fermions,
is close to the value we find here $S=L \ln 2$.

For the quantum dot regime
a direct comparison of numerical $n_k$ values with
the theoretical Fermi-Dirac distribution (\ref{eq3})
is possible as it is shown in Fig.~\ref{fig2}
and Fig.~\ref{fig5}. 
A determination of $n_k$ values 
hidden quasi-particles in
the SYK regime remains an open problem. 
We note that in a certain sense
the thermodynamic computations
done in \cite{hanada,garcia} assume the thermal
distribution over many-body levels $E_m$ 
produced by a certain thermostat
with temperature $T$ 
being in contact with the quantum dot or the SYK black hole.
We think that a bath thermostat is not realistic for the case
of black holes which are well isolated objects
(at least in a first approximation).
At the same time the DTC is well defined for 
isolated systems.

\section{Discussion}

We demonstrate that the dynamical thermalization 
takes place for interacting fermions 
in a quantum dot for interactions
being above a certain threshold. We argue that this threshold
is given by the  {\AA}berg criterion  (\ref{eq2}) \cite{aberg1,aberg2,jacquod}.
Above the threshold we directly show that the DTC is valid and
the eigenstates
of the many-body Hamiltonian (\ref{eq2})
are described by the Fermi-Dirac thermal 
distribution (\ref{eq3}) over 
one-particle orbitals  in the quantum dot regime ($g=V/J > 1$).
For the SYK black hole regime
with the DTC we reproduce
the maximal entropy values and
argue that hidden quasi-particle states
reproduce dependence of entropy $S$ on energy $E$.
We show  that the verification of the DTC validity 
is done in a most optimal way
by the comparison of the numerically obtained entropy on energy
dependence $S(E)$ with the theoretical Fermi-Dirac distribution
(\ref{eq3}) in the quantum dot regime and 
in the SYK black hole case where the semi-empirical
quasi-particle basis is still to be found.
We point that in previous studies
\cite{hanada,garcia} the thermodinamical characteristics have been
obtained in assumption of a contact between the system and external
bath thermostat that is opposite to the dynamical thermalization
concept.

\begin{figure}[h]
\begin{center}
\includegraphics*[width=8.2cm]{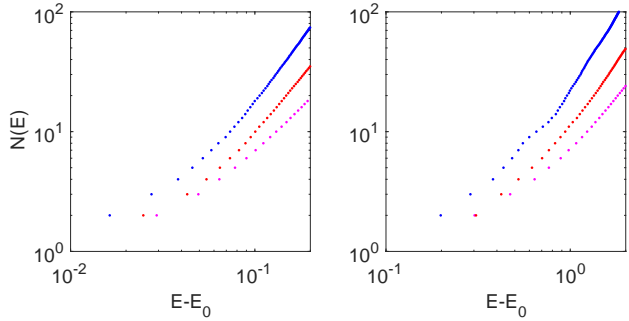}
\end{center}
\vglue -0.3cm
\caption{(Color on-line) 
Dependence of integrated number of states
$N(E)$ from ground state $E_0$ 
up to energy $E$  for the SYK black hole regime
at $V=0$
(left panel) and the quantum dot regime 
$V = \sqrt{14}$ (right panel).
Here 
$M=14, L=6, N=3003$ with average over $N_r=10$ 
disorder realizations (blue symbols), 
$M=12, L=5, N=792, N_r=38$ (red symbols),
$M=10, L=4, N=210, N_r=150$ (magenta symbols); in all cases $J=1$.
Only a vicinity of ground state energy is shown.
}  
\label{fig7}
\end{figure}

The extension of the SYK model ($g \ll 1$) to the quantum dot regime
($g>1$) described by the Hamiltonian (\ref{eq2}) 
rises several new questions. Indeed, excitations
at $g >1$ have an energy gap $\Delta E \propto 1/L^{3/2}$
which drops only algebraically with the number of fermions $L$.
In contrast it is expected that the low energy
excitations at $g \ll 1$ have excitation energy 
$\delta E \propto \exp(- C L)$ which drops exponentially
with $L$ ($C$ is some constant). Indeed, 
the results of Fig.~\ref{fig7}, showing
the average integrated number of states $N(E)$ 
above the ground energy $E_0$, confirm that the 
low excitation energies are by an order of magnitude smaller
for $g \ll 1$ compared to the case of $g > 1$.
However, much larger values of $L$ and 
better averaging over many disorder
realizations $N_r$ are required to distinguish
firmly algebraic and exponential dependencies on 
number of fermions. At the same time the numerical
results for the SYK model with Majorana fermions
confirms the exponential drop of $\delta E$ with $L$
for $g \ll 1$ \cite{garcia} 
while the Landau theory of Fermi liquid
guaranties an algebraic drop of $\delta E$ with $L$
for $g \gg 1$.
We expect that a quantum phase transition can take
place between these two regimes at a certain
critical conductance $g_c$ of a quantum dot.
The interesting question on interpretation of negative
temperatures for the SYK black holes remains for further studies.

Our results show that quantum dots
with moderate conductance values $g \sim 1$
can be close for the SYK black hole regime.
Thus an experimental investigations of such
quantum dots can open new perspectives for
studies of the SYK model.
Such solid state systems with $g \ll 1$ 
have been already studied with 2D lattice
of coupled Sinai billiards \cite{kvon}.
Another possibility to investigate strongly 
interacting fermions is to consider the regime
of Anderson localization where
the conductance takes values $g \sim 1$
inside Thouless blocks and interactions
are strong near the Fermi level \cite{thouless,akkermans,tip}.

Finally extending the proposal of SYK modeling with cold atoms
\cite{hanada} we propose to consider a case
of  Wigner crystal in a periodic potential
which, as SYK,  is characterized by exponentially small
energy excitation inside the pinned Aubry phase 
\cite{fki}. Such an Aubry phase has been recently realized
with cold ions in optical lattices \cite{vuletic}.

We hope that our results will stimulate further research of duality
between SYK black holes and  
quantum dots with strongly interacting fermions.
We note that the dynamical thermalization concept
is especially interesting for black holes
which can be naturally considered as isolated objects
without any contact with thermostat.

This work was done in the frame of LABEX NEXT project THETRACOM.

\end{document}